\documentclass[final]
  {aipproc}
\newcommand{\be}{\begin{eqnarray}}
\newcommand{\ee}{\end{eqnarray}}

\newcommand{\ave}[1]{\left\langle #1 \right\rangle}

\layoutstyle{6x9}

\begin{document}

\title{Particle multiplicities and fluctuations\\ in 200 GeV Au-Au collisions}


\author{Giorgio Torrieri}{ address={Department of Physics, 
McGill University, Montreal, QC H3A-2T8, Canada}
}

\author{Sangyong Jeon}{
  address={Department of Physics, McGill University, Montreal, QC H3A-2T8, Canada}
,altaddress={RIKEN-BNL Research Center, Upton NY, 11973, USA}  additional visiting address}

\author{Johann Rafelski}{
  address={Department of Physics, University of Arizona, Tucson AZ 85721, USA}
}

\begin{abstract}
We use the statistical hadronization model (SHM)
 to describe hadron multiplicity yields and fluctuations. We consider
 200 GeV Au-Au collisions, and
 show that both event averaged yields  of stable particles
 and resonances, and event-by-event fluctuation of the $K/\pi$ 
ratio can be described within the SHM using the same set of thermal parameters, provided that the phase space occupancy parameter 
value is   significantly above chemical equilibrium, 
and the freeze-out temperature is  $\sim 140$ MeV.   We
present predictions that allow to test the consistency
of our results.
\end{abstract}

\maketitle


The statistical hadronization model (SHM) 
\cite{Fer50,Pom51,Lan53} has been extensively
 applied to the study of soft particle production
 in hadronic systems.   When it includes the full spectrum of hadronic
resonances \cite{Hag65}, the SHM,  with judiciously chosen 
(fitted) parameters,  quantitatively describes
the abundances of all   hadrons produced 
in heavy ion collisions at all considered reaction energies
\cite{gammaq_energy}.   

The ability of the SHM to describe not
 just averages, but event-by-event multiplicity fluctuations
has  not been widely investigated.
 Event-by-event fluctuations have attracted
theoretical~\cite{fluct2,jeonkochratios,fluct4,fluct6}
and experimental interest, 
both as a consistency check
for existing models ~\cite{fluct2,jeonkochratios} and as a way 
to search  for new physics \cite{fluct4}.

The objective of this work is to determine weather the SHM 
can describe both yields and fluctuations with the same 
parameters. We obtain a good fit to  200 GeV RHIC 
experimental data including both yields and fluctuations measurements, discuss the results 
in the context of the bulk properties of the matter created at RHIC, 
and present  predictions allowing further tests of the model.   

The statistical hadronization model assumes that final states 
are produced in proportion to their phase
space size.   The first and second cumulants of this probability 
distribution give, respectively, the average value over all events its event-by-event fluctuation.

In this work we use the Grand Canonical (GC) 
 ensemble, as implemented in open-source software \cite{share}, to calculate fluctuations and yields.
We motivate this choice by the fact that the considered RHIC experiments
observe the mid-rapidity slice of the system, comprising roughly 1/8 of the total 
hadron multiplicity.  The further 
assumption of (approximate) boost invariance 
at mid-rapidity allows to image   this rapidity 
slice into  a domain in configuration space.
The matter content in this space domain is expected 
to   be in contact and exchanging energy and conserved quantum numbers 
 with the unobserved regions. This than creates the 
GC system we consider on an event-by-event basis.

If the freeze-out temperature throughout the system is 
the same, a rather simple model   can be used to obtain both 
yields and fluctuations \cite{cleymans,prlfluct}. However, one should 
note that even if practically all produced particles originate
in statistical model processes, their fluctuations could comprise 
novel creation mechanisms, related to their formation dynamics. 
However, such novel mechanisms are not present in all observables, and 
the observables  we consider here seem to follow non-equilibrium SHM calculations. 

The final state particle yield can then be computed as a function of
the particle properties, resonance decay tree, freeze-out temperature and fugacities (technical details are found  
 in our recent  report~\cite{qm2005nukleonika}).  
While the temperature controls the particle yield 
dependence on the mass, the fugacity $\lambda$ describes 
both  the yield of conserved quantities (such as baryon number, charge and strangeness) across all 
particles, and  the absolute  yields which depend on the degree of chemical 
equilibration. It is common to introduce chemical 
potentials associated with each conservation law,  $\mu=T\ln \lambda$, while the 
fugacities associated with the chemical nonequilibrium condition are called $\gamma$. 
 
Detailed balance requires that the particle 
fugacity be given by the conserved charge fugacity 
to the power of the particle's `charge', generalized to contain all
conserved quantities (electrical charge, baryon number etc.) .
Thus, for a particle with $q, (\overline{q})$ light (anti)quarks, 
$s, (\overline{s})$ strange (anti)quarks and isospin $I_3$ we have
\begin{eqnarray}
\label{chemneq}
\lambda_i^{\mathrm{eq}}=
\lambda_{q}^{q-\overline{q}} \lambda_{s}^{s -\overline{s}}
\lambda_{I_3}^{I_3} 
\end{eqnarray}
 However, the condition of chemical equilibrium might no longer hold  
when the fireball  is rapidly  expanding.
Thus, chemical parameters  acquire a {\em kinetic}
 (time-dependent) component parametrized in terms 
of phase space occupancies, 
\begin{eqnarray}
\label{chemneq2}
\lambda_i = \lambda_i^{\mathrm{eq}}
\gamma_q^{q+\overline{q}} \gamma_s^{s+\overline{s}},
\end{eqnarray}
where $\gamma_q = 1,\gamma_s = 1$ at equilibrium.
Even in chemical nonequilibrium 
the particle fugacity $\lambda_i$ is the parameter 
controlling the  particle yield, and the first and 
second cumulants can be calculated from the partition 
function in the usual way \cite{jansbook,chemistrywebsite}. 

If the expanding system undergoes a fast phase transition 
 from a Quark Gluon Plasma (QGP) to a hadron gas (HG),
 chemical non-equilibrium  \cite{sudden} and 
super-cooling \cite{csorgo} are expected to arise 
given the requirement that  entropy has to increase
while the transition occurs from a high to low 
entropy density phase.  The virtue of a
hadronization temperature near $140$ MeV, 
and an over-saturated phase space
 ($\gamma_q \sim 1.5,\gamma_s \sim 2$), 
 is a match of both the energy and entropy density 
 between the QGP and HG phases.

Fits to experimental data  at both SPS
and RHIC energies  support these values
 of $\gamma_{q,s}$.  Moreover, best fit
 $\gamma_{q,s}>1$ arises for a critical 
reaction energy \cite{gammaq_energy}  
(corresponding to the energy of the $K/\pi$ 
``horn'' \cite{horn}) and system size \cite{gammaq_size}, 
as expected from the interpretation of $\gamma_q$ 
as a manifestation of a phase transition. 
 However, even though the fits performed in \cite{gammaq_energy} 
strongly favor the chemical non-equilibrium, they 
do not  rule out  equilibrium models.
The equilibrium model remains marginally   compatible 
with data giving a less convincing but statistically 
still non-absurd validity.This happens 
since in a fit considering only particle  yields, the 
chemical non-equilibrium phase space occupancies
$\gamma_s$ and $\gamma_q$ correlate with freeze-out 
temperature  \cite{gammaq_energy}, making a distinction between
 a higher temperature equilibrated freeze-out 
($\gamma_q=1,\gamma_s\leq 1$) scenario and 
a supercooled scenario where $\gamma_{q,s}>1$ difficult.  
When full chemical nonequilibrium is allowed for, $\gamma_q\simeq 1.6$ and 
$T=140$ MeV is found.  The best fit freeze-out temperature when 
full chemical equilibrium
 is assumed varies between studies, ranging from $T=155$ MeV in 
latest SHARE based  studies \cite{gammaq_energy},  $T=165$ MeV for those 
carried out 2 years ago by STAR experimental group \cite{barannikova},
 to $T=177$   MeV offered in the initial RHIC data exploration in which 
strange particles were not yet fully incorporated \cite{bdm}.  

 The study presented  in  \cite{qm2005nukleonika}
 makes it clear that the dependence of the fluctuation
\be
\sigma_X^2 = \frac{\ave{X^2}-\ave{X}^2}{\ave{X}}
\label{fluctdef}
\ee
 on $T$ and $\gamma_q$
 is different, allowing us to independently determine  these two variables. 
 A higher temperature decreases the fluctuations with respect to 
 the Poisson value $\sigma=1$, expected for a  Boltzmann distribution,
since it introduces greater particle  correlations arising from  increased
resonance decay contribution.  Increasing $\gamma_q$   rapidly increases
 fluctuations of quantities related to pions, due to the fact
that at $\gamma_q>1$ $\lambda_{\pi}$ rapidly approaches $e^{m_{\pi}/T}$, 
giving fluctuations an extra increase compared to  
 yields \cite{prlfluct,qm2005nukleonika}.

By virtue of the implied physical picture, equilibrium 
models generally assume a long time span
between chemical (particle production) and thermal (particle
scattering) freeze-out, which would alter considerably 
the multiplicity of directly detectable resonances.  
In the chemical non-equilibrium supercooled freeze-out picture, however, it is natural to assume that particle scattering 
after emission is negligible \cite{sudden} and thus
one can in most cases assume that the thermal and chemical
freeze-out temperatures are the same. Hence,
a reliable way to probe the 
re-interaction period would be instrumental for our understanding of how 
the fireball produced in heavy ion collisions breaks up. 

We have recently shown \cite{prlfluct} that a comparison of  fluctuations to directly
 detected resonances probes the interval between chemical and thermal freeze-out.
Consider, for example, $\sigma_{K^+/\pi^-}$.
The numerator and the denominator terms in this ratio are 
linked by a large correlation term due to the 
$K^{*0}(892) \rightarrow K^+ \pi^-$ decay.   
This correlation probes the $K^{*0}(892)$ 
abundance at the initial {\it chemical } freeze-out, since subsequent rescattering of $K^* \rightarrow K \pi$ decay products 
{\em or} on-shell $K^*$ regeneration from in-medium $K^+ \pi^-$ pairs
does not alter the final abundance of  $\pi^+$ and a $K^-$.
On the other hand, a direct measurement of the $K^{*0}(892)/K^-$ ratio through invariant mass reconstruction measures the $K^{*0}(892)$ abundance at {\em thermal} freeze-out, 
after all rescattering/regeneration ceased. 
Hence, comparing the $K^+/\pi^-$ fluctuation   to the  $K^{*0}(892)/K^-$ 
ratio provides a gauge for the effect of the hadronic reinteraction period 
on particle abundances. A strong constraint arises in a   
model where chemical and thermal freeze-out coincide, as   both 
observables must be described with the same set of statistical parameters
determined by global yields of other particles. In this way one 
can argue that the $K^+/\pi^-$ fluctuation  and the  $K^{*0}(892)/K^-$ 
relative yield measured by invariant mass method offer a decisive
test of the chemical non-equilibrium sudden hadronization reaction picture.

While they are phenomenologically powerful, fluctuation measurements 
are also vulnerable to systematic effects which need to be carefully considered.
Volume and centrality fluctuations, difficult to describe in a model-independent way, 
can be taken care of by considering event by event fluctuations of particle ratios, 
where the fluctuation in volume cancels out, event by event, to first order.     
This leaves, however, possibly large effects due to limited experimental acceptance.
These can usually be subtracted by considering fluctuations measured within fake events, 
created using mixed event techniques \cite{pruneau}.
Such ``static'' fluctuation, with, by definition, no correlations between particles, 
can be described by a purely Poisson term, where fluctuation is governed by particle yields:
\be
\sigma_{stat}^2 = \frac{1}{\ave{N_1}}+\frac{1}{\ave{N_2}}
\label{poissrat}
\ee
It can also be seen that mixed-events, made from tracks measured in a given detector, 
also contain the effect due to that detector's acceptance.

Subtracting $\sigma_{stat}$ from the total fluctuation leaves the ``dynamical'' 
fluctuation term:
\be
\label{dynrat}
\sigma^{dyn}=\sqrt{\sigma^2-\sigma_{stat}^2},
\ee
which comprises the   physically interesting effects  such as
resonance decay correlations, Bose-Einstein correlations,   and 
eventual new dynamics. 
Provided certain assumptions for the detector response function hold 
(see appendix A of \cite{pruneau}), this dynamical fluctuation is a ``robust'', detector-independent observable.

When using measured $\sigma^{dyn}$s in fits to experimental data, 
the data-sample should include $\sigma^{dyn}$, particle yields 
(which are needed to determine the Poisson contribution 
to $\sigma_{stat}$ as per Eq.\,(\ref{poissrat})) and particle ratios.
We have performed a fit incorporating all STAR ratios given in Ref.\,\cite{barannikova}, 
with the exception of the $\Delta^{++}/p$, which the STAR collaboration has since begun 
to reevaluate, and we also for the present ignore the
$\overline{\Omega}/\Omega>1$, which cannot be fitted 
with the SHM \cite{qm2005nukleonika,liu}. On the other hand, 
we have included in the fit procedure the preliminary value
of $\sigma^{dyn}_{K/\pi}$ measured by STAR \cite{supriya}, 
as well as the published yield for $\phi$ \cite{starphi200} 
and $\pi^-$ \cite{starpi200}.
We assume \cite{starpi200,barannikova_private} full feed-down correction 
for $K_{S,L} \rightarrow \pi^{\pm}$ and  $\Lambda \rightarrow \pi$ 
weak decays, and no correction for $\Lambda \rightarrow p$.

\begin{figure}[tb]
\includegraphics[width=7.6cm]{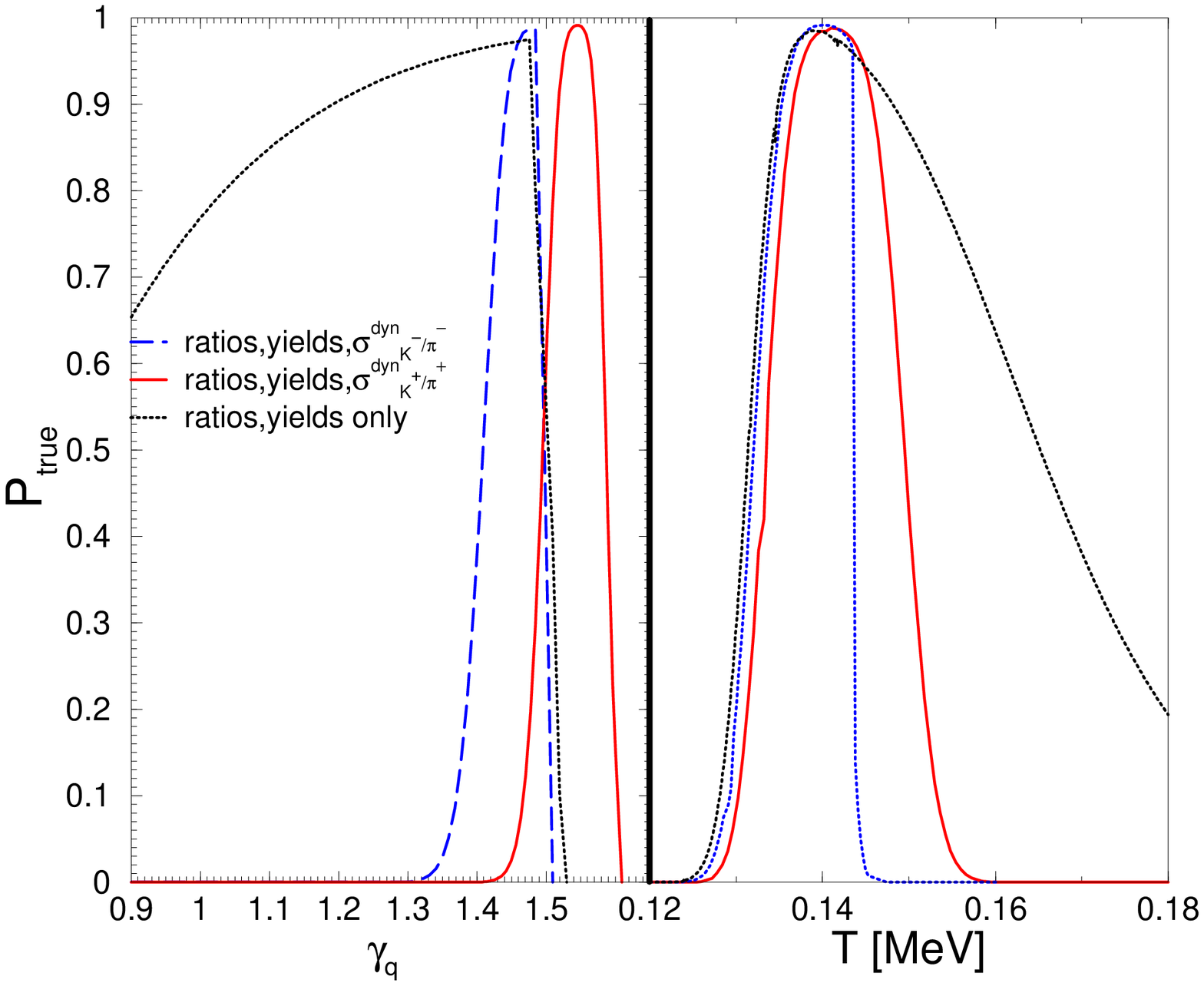}
\hspace*{0.6cm}
\includegraphics[width=7.6cm]{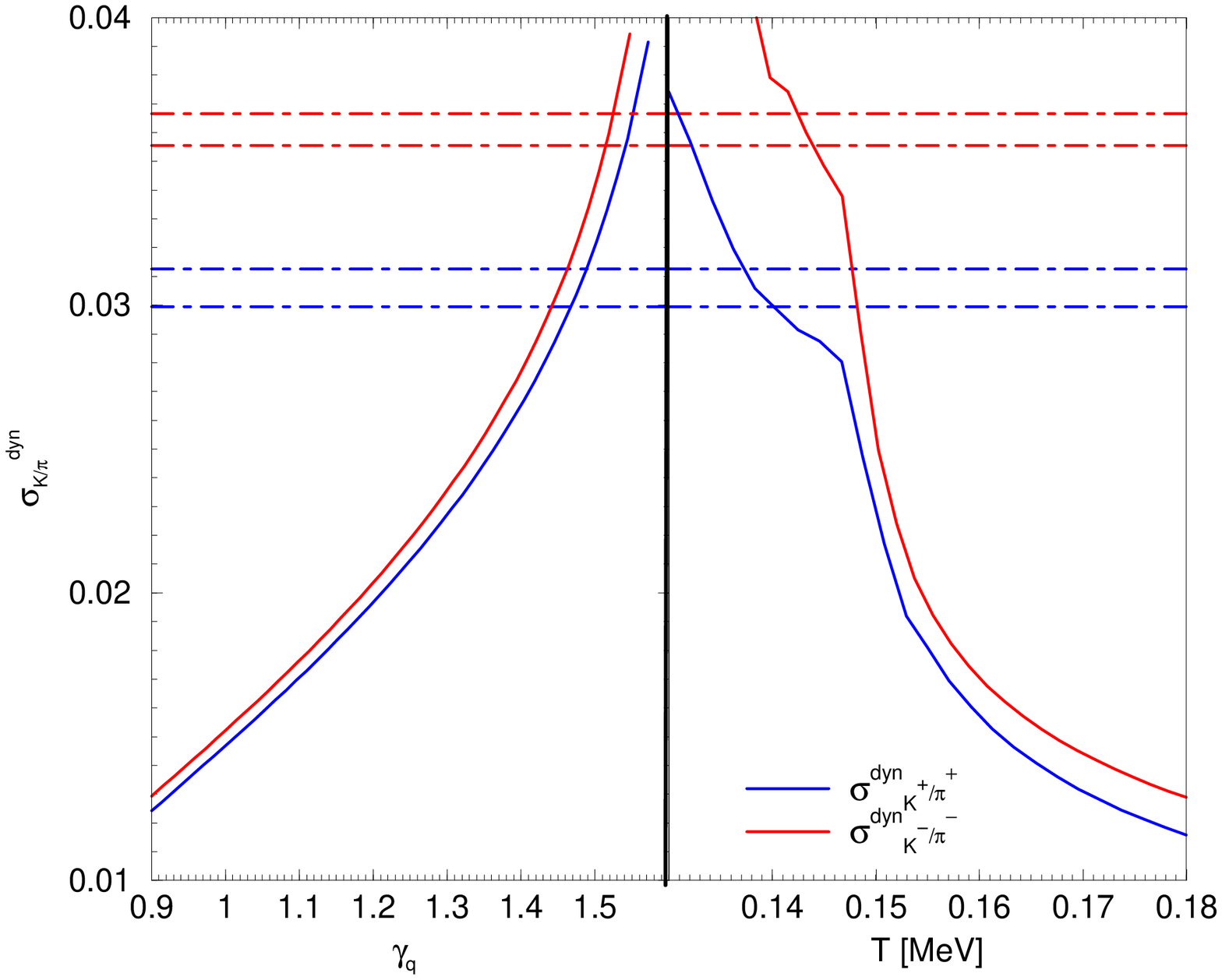}
\caption{
\label{prof_200}
Left:   statistical significance ($P_{true}$) profiles for freeze-out Temperature $T$ 
and light quark phase space occupancy $\gamma_q$  obtained in  the fit shown 
in Fig. \ref{graph_200}. 
Right:  Sensitivity of $\sigma_{K/\pi}^{dyn}$ to freeze-out temperature 
and $\gamma_q$ for the fit in Fig. \ref{graph_200}.    
At each point in the abscissa, a fit is performed , with only the particle 
yield data-points used, varying all fit parameters except the one on the abscissa.
$\sigma_{K/\pi}^{dyn}$ is evaluated using the best fit parameters.  The dot-dashed lines 
refer to the experimental limits \cite{supriya}  for the $\sigma_{K^+/\pi^+}$ (blue, lower values) 
and $\sigma_{K^-/\pi^-}$ (red, higher values).
  See  \cite{qm2005nukleonika} for further details. }
\end{figure}

The fit parameters include the overall normalization, the
freeze-out temperature $T$, $\lambda_{q,s,I_3}$ and $\gamma_{q,s}$.
We also require, by implementing them as ``data-points'', strangeness, electrical 
charge and baryon number conservation:
$ \ave{s-\overline{s}}=0 \phantom{AA}\ave{Q}/\ave{B}=(\ave{Q}/\ave{B})_{Au}=0.4$.  
The fit's statistical significance ($P_{true}$) profile is shown in the left panel 
of Fig. \ref{prof_200}. To obtain a profile for $P_{true}$ a fit was performed for   each 
fixed parameter point on the abscissa, all fit parameters except the one shown 
on the abscissa were varied.  It can be seen that the fit tightly constrains $\gamma_q$ well above
the equilibrium value,  accompanied by $T\sim 140$ MeV, in good agreement 
of the prediction of the supercooled hadronization scenario. 

The right panel of Fig. \ref{prof_200} shows the sensitivity 
of $\sigma_{K/\pi}^{dyn}$ to $\gamma_q$ and temperature, 
and explains why the correlation between $T$ and $\gamma_q$ 
disappears when fluctuations are taken into account.  
As can be seen, a fit assuming the  chemical equilibrium ($\gamma_q=1,T=155$ MeV \cite{gammaq_energy}).
 misses $\sigma_{K/\pi}^{dyn}$ by many standard deviations. On the contrary
the chemical nonequilibrium fit seems to be right where this fluctuation 
is measured.    Introducing exact conservation for strangeness 
within the observed window (canonical ensemble) would decrease the theoretical 
$\sigma_{K}$ \cite{nogc2}, thereby increasing the  chemical equilibrium  theory to 
experiment discrepancy.  It is only through $\gamma_q>1$ that $\sigma_{K/\pi}^{dyn}$ 
increases to the point where it becomes compatible with the experimental value.  

\begin{figure*}[tb]
\includegraphics[width=7.cm]{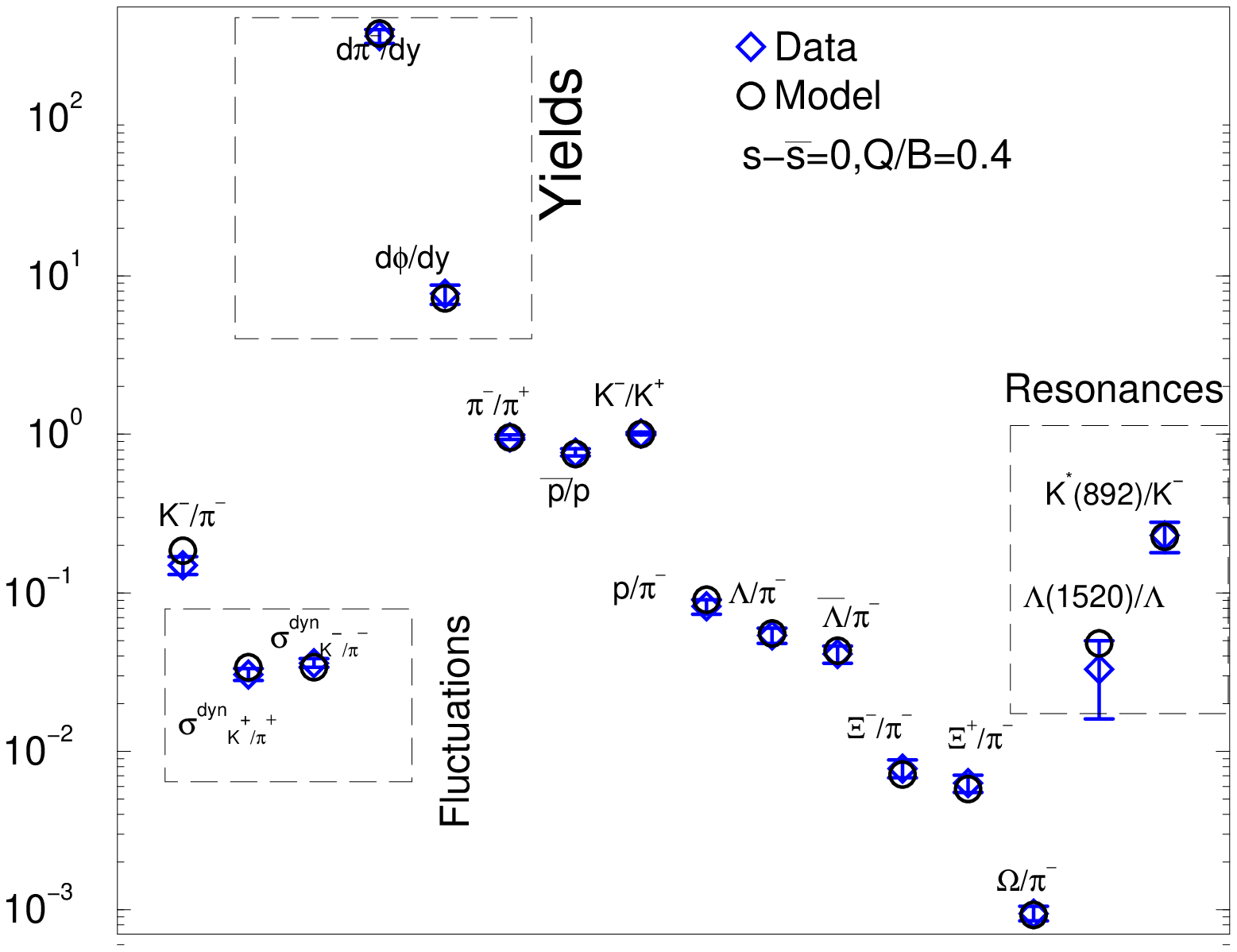}
\includegraphics[height=5.6cm]{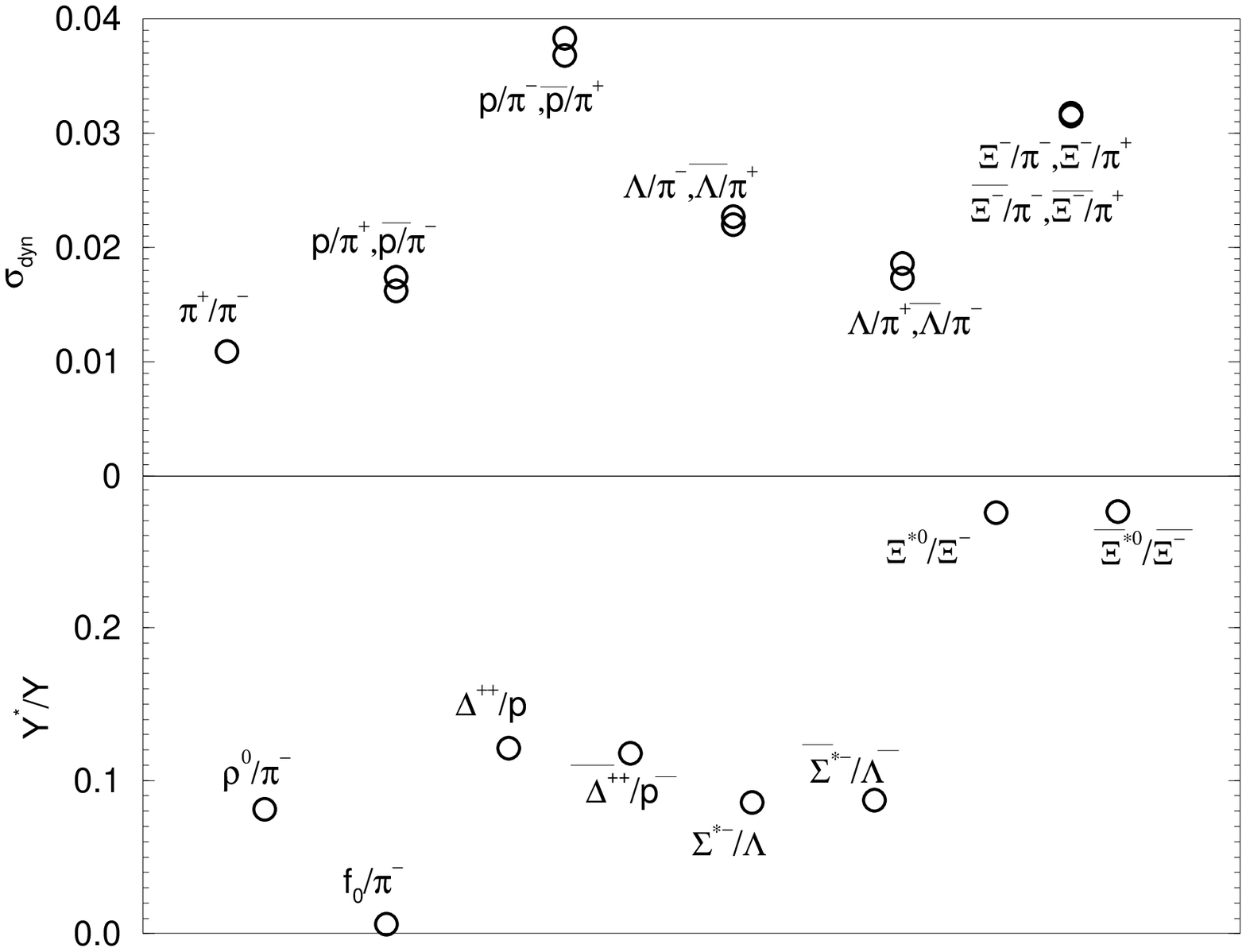}
\caption{\label{graph_200}
Left: Fit of preliminary 200 GeV data, including the $K^{\pm}/\pi^{\pm}$ 
fluctuations and the $K^* (892)$ and $\Lambda^* (1520)$ resonance
Right: Predictions of resonances and event-by-event 
fluctuations of their decay product ratios with the best fit parameters.}
\end{figure*}

An eye assessment of the fit's goodness is provided
by the left panel of Fig. \ref{graph_200}.    
As can be seen, the fit gives an adequate description 
of all particle yields, including the
resonance $(K^{*0}(892)+\overline{K^{*0}}(892))/K^-$ and 
$\Lambda(1520)/\Lambda$.   It can also adequately describe 
the event-by-event fluctuations of   $K^+/\pi^+$ \textit{or} 
$K^-/\pi^-$. 

$K^{\pm}/\pi^{\pm}$ fluctuations do not directly test the simultaneous freeze-out hypothesis, since
$K^- \pi^-$ and $K^+ \pi^+$ are not correlated by resonances.  To test sudden freeze-out, we have used the best fit parameters to predict 
the yields of several resonances subject to current experimental investigation 
($\rho^0$,$f_0(950)$,$\Delta^{++}$,$\Sigma^{*+}(1385)$,$\Xi^*(1530)$) 
as well as the dynamical fluctuation of the ratio of their decay products
($\pi^+/\pi^-$,$p/\pi^{\pm}$,$\Lambda/\pi^{\pm}$,$\Xi/\pi^{\pm}$).   
The result is shown in Fig \ref{graph_200}, right panel.

Note the significant difference between
ratios such as $p/\pi^+$ and $p/\pi^-$, 
while the fluctuations of $p/\pi^+$ 
and $\overline{p}/\pi^-$ are substantially identical.   
This systematics, which repeats itself
when the $\Lambda/\pi$ ratio is considered, 
is due to the correlations provided by the leading 
resonance decay ($\Delta \rightarrow p \pi$, 
$\Sigma^* (1385) \rightarrow \Lambda \pi$).
Thus, the combined measurement of the resonance 
and the ratio of decay products yields a very powerful constraint
on the simultaneous freeze-out model considered here, 
and it will be interesting to see to what extent will 
the model agree with data.    
In particular, the difference between $\sigma_{K^+/\pi^+}^{dyn}$ 
and $\sigma_{K^-/\pi^-}^{dyn}$ is intriguing, 
since the isospin chemical potential required to reproduce it 
($\lambda_{I3} \sim 0.96$) is excluded through ratios
such as the $\pi^+/\pi^-$.  
It remains to be seen weather this result is
 due to experimental systematics not analyzed 
within the preliminary measurement, or weather 
additional theoretical insights are needed to describe it.

In conclusion, we have shown that the SHM can describe both the yields 
and event-by-event fluctuations measured so far in RHIC 200 GeV Au-Au collisions, 
provided that phase space is saturated above equilibrium and the system 
is super-cooled with respect to the phase transition temperature.  
We have justified this scenario in the context of a 
fast phase transition from a high-entropy
phase, and argued that the simultaneous description 
of yields and fluctuations is consistent with an explosive 
freeze-out, where interactions after hadronization are negligible.
We have predicted experimental observables suitable for testing this 
model further, and eagerly await more published data to determine 
to what extent can the SHM
 account for both yields and fluctuations in light and strange 
hadrons produced in heavy ion collisions.

\begin{theacknowledgments}
Work supported in part by grants from
the U.S. Department of Energy  (J.R. by DE-FG02-04ER41318)
the Natural Sciences and Engineering research
council of Canada, the Fonds Nature et Technologies of Quebec.
G. T. thanks the Tomlinson foundation  for support.
S.J.~thanks RIKEN BNL Center
and U.S. Department of Energy [DE-AC02-98CH10886] for
providing facilities essential for the completion of this work.
\end{theacknowledgments}


\begin{thebibliography}
\expandafter\ifx\csname natexlab\endcsname\relax\def\natexlab#1{#1}\fi
\expandafter\ifx\csname bibnamefont\endcsname\relax
  \def\bibnamefont#1{#1}\fi
\expandafter\ifx\csname bibfnamefont\endcsname\relax
  \def\bibfnamefont#1{#1}\fi
\expandafter\ifx\csname citenamefont\endcsname\relax
  \def\citenamefont#1{#1}\fi
\expandafter\ifx\csname url\endcsname\relax
  \def\url#1{\texttt{#1}}\fi
\expandafter\ifx\csname urlprefix\endcsname\relax\def\urlprefix{URL }\fi
\providecommand{\bibinfo}[2]{#2}
\providecommand{\eprint}[2][]{\url{#2}}


\bibitem{Fer50}
E. Fermi, {Prog. Theor. Phys.} {\bf 5}, 570 (1950).

\bibitem{Pom51}
I. Pomeranchuk, {Proc. USSR Academy of Sciences} (in Russian)
{\bf 43}, 889 (1951). 

\bibitem{Lan53}
  L.~D.~Landau,
  Izv.\ Akad.\ Nauk Ser.\ Fiz.\  {\bf 17} (1953) 51.

\bibitem{Hag65}
R. Hagedorn, {Suppl. Nuovo Cimento}  {\bf 2}, 147 (1965).


\bibitem{gammaq_energy}
  J.~Letessier and J.~Rafelski,
  arXiv:nucl-th/0504028.

\bibitem{fluct2}
S.~Jeon, V.~Koch, K.~Redlich and X.~N.~Wang,
Nucl.\ Phys.\ A {\bf 697}, 546 (2002)

\bibitem{jeonkochratios}
S.~Jeon and V.~Koch,
Phys.\ Rev.\ Lett.\  {\bf 83}, 5435 (1999)

\bibitem{fluct4}
S.~Jeon and V.~Koch,
Phys.\ Rev.\ Lett.\  {\bf 85}, 2076 (2000)

\bibitem{fluct6}
  C.~Pruneau, S.~Gavin and S.~Voloshin,
  Phys.\ Rev.\ C {\bf 66}, 044904 (2002)
  [arXiv:nucl-ex/0204011].

\bibitem{share}
G.~Torrieri,  {\it et al.},
Comm. in Computer physics in press arXiv:nucl-th/0404083

\bibitem{cleymans}
  J.~Cleymans, K.~Redlich,
  Phys.\ Rev.\ C {\bf 60}, 054908 (1999).


\bibitem{prlfluct}
  G. Torrieri, S. Jeon, J. Rafelski
  Submitted to PRL
  [arXiv:nucl-th/0503026].

\bibitem{qm2005nukleonika}
  G.~Torrieri, S.~Jeon and J.~Rafelski,
  arXiv:nucl-th/0509067.

\bibitem{jansbook}
J.~Letessier and J.~Rafelski,
Cambridge Monogr.\ Part.\ Phys.\ Nucl.\ Phys.\ Cosmol.\  {\bf 18}, 1 (2002),

\bibitem{chemistrywebsite}
Material usually presented in   physical chemistry textbooks, see for example:\\
{\it http://www2.mcdaniel.edu/Chemistry/ch307.notes/Chemical$\%$20Equilibrium.html}




\bibitem{sudden}
J. Rafelski and J. Letessier,
{\it Phys. Rev. Lett.}  {\bf 85}, 4695 (2000).

\bibitem{csorgo}
T. Csorgo and L. Csernai,
arXiv:hep-ph/9312330

\bibitem{horn}
 D.~Flierl {\it et al.}  [NA49 Collaboration],
  arXiv:nucl-ex/0410041.

\bibitem{gammaq_size}
  J.~Rafelski, J.~Letessier and G.~Torrieri,
  Phys.\ Rev.\ C {\bf 72}, 024905 (2005)
  [arXiv:nucl-th/0412072].

 \bibitem{barannikova}
O.~Barannikova  [STAR Collaboration],
arXiv:nucl-ex/0403014  and references therein.

\bibitem{bdm}
  P.~Braun-Munzinger, D.~Magestro, K.~Redlich and J.~Stachel,
  Phys.\ Lett.\ B {\bf 518}, 41 (2001).

\bibitem{supriya}
  S.~Das  [STAR Collaboration],
  arXiv:nucl-ex/0503023.



\bibitem{pruneau}
  C.~Pruneau, S.~Gavin and S.~Voloshin,
  Phys.\ Rev.\ C {\bf 66}, 044904 (2002)
  [arXiv:nucl-ex/0204011].


\bibitem{liu}
M.~Bleicher {\it et al.},
Phys.\ Rev.\ Lett.\  {\bf 88}, 202501 (2002)
[arXiv:hep-ph/0111187].

\bibitem{starphi200}
J.~Adams {\it et al.}  [STAR Collaboration],
Phy. Lett. B in press,
[arXiv:nucl-ex/0406003].

\bibitem{starpi200}
J.~Adams {\it et al.}  [STAR Collaboration],
Phys.\ Rev.\ Lett.\  {\bf 92}, 112301 (2004)
[arXiv:nucl-ex/0310004].

 \bibitem{barannikova_private}
O.~Barannikova  [STAR Collaboration], private communication

\bibitem{nogc2}
  V.~V.~Begun, M.~Gazdzicki, M.~I.~Gorenstein and O.~S.~Zozulya,
  Phys.\ Rev.\ C {\bf 70}, 034901 (2004)
  [Phys.\ Rev.\ C {\bf 72}, 014902 (2005)]
  [arXiv:nucl-th/0404056].

\end{thebibliography}
\end{document}